\documentclass[10pt,superscriptaddress,aps,prl,twocolumn,showpacs,nofootinbib, longbibliography, notitlepage,floatfix]{revtex4-2}
\usepackage{mathtools}
\usepackage[english]{babel}
\usepackage{booktabs}
\usepackage{amsbsy}
\usepackage{amssymb}
\usepackage{amstext}
\usepackage{bm}
\usepackage{xcolor}
\usepackage{array}
\usepackage{booktabs}
\usepackage{colortbl}
\usepackage{xcolor}
\usepackage{amssymb}
\usepackage{textcomp}

\usepackage{multirow}

\definecolor{darkblue}{rgb}{0, 0, 1}
\newcommand{\RN}[1]{%
	\textup{\uppercase\expandafter{\romannumeral#1}}%
}
\usepackage[colorlinks=true,citecolor=blue,urlcolor=blue]{hyperref}

\makeatletter

\@ifundefined{textcolor}{}
{%
	\definecolor{BLACK}{gray}{0}
	\definecolor{WHITE}{gray}{1}
	\definecolor{RED}{rgb}{1,0,0}
	\definecolor{GREEN}{rgb}{0,1,0}
	\definecolor{BLUE}{rgb}{0,0,1}
	\definecolor{CYAN}{cmyk}{1,0,0,0}
	\definecolor{MAGENTA}{cmyk}{0,1,0,0}
	\definecolor{YELLOW}{cmyk}{0,0,1,0}
}

\newcommand{\beq}{\begin{equation}}
\newcommand{\eeq}{\end{equation}}
\newcommand{\beqa}{\begin{eqnarray}}
\newcommand{\eeqa}{\end{eqnarray}}

\newcommand{\ket}[1]{\vert #1\rangle}

\usepackage{amsmath}
\usepackage{graphicx}
\usepackage{amssymb}
\usepackage{txfonts,color}
\makeatother

\begin{document}


\title{Optimal Control for Open Quantum System in Circuit Quantum Electrodynamics}

\author{Mo Zhou}
\affiliation{Institute for Quantum Science and Technology, Department of Physics, Shanghai University, Shanghai 200444, China}
\affiliation{Instituto de Ciencia de Materiales de Madrid (CSIC),
Cantoblanco, E-28049 Madrid, Spain}

\author{F. A. C\'ardenas-L\'opez}
\email{f.cardeans.lopez@fz-juelich.de}
\affiliation{Forschungszentrum J\"ulich GmbH, Peter Gr\"unberg Institute, Quantum Control (PGI-8), 52425 J\"ulich, Germany}

\author{Sugny Dominique}
\affiliation{ Laboratoire Interdisciplinaire Carnot de Bourgogne, CNRS UMR 6303, Université de Bourgogne, BP, 47870, F-21078 Dijon, France}

\author{Xi Chen}
\email{xi.chen@csic.es}
\affiliation{Instituto de Ciencia de Materiales de Madrid (CSIC), Cantoblanco, E-28049 Madrid, Spain}


\date{\today}

\begin{abstract}
We propose a quantum optimal control framework based on the Pontryagin Maximum Principle to design energy- and time-efficient pulses for open quantum systems. By formulating the Langevin equation of a dissipative LC circuit as a linear control problem, we derive optimized pulses with exponential scaling in energy cost, outperforming conventional shortcut-to-adiabaticity methods such as counter-diabatic driving. When applied to a resonator dispersively coupled to a qubit, these optimized pulses achieve an excellent signal-to-noise ratio comparable to longitudinal coupling schemes across varying critical photon numbers. Our results  provide a significant step toward efficient control in dissipative open systems and improved qubit readout in circuit quantum electrodynamics.
\end{abstract}

\maketitle



\textit{Introduction.--}
Controlling quantum systems is fundamental for advancing quantum technologies, with quantum optimal control serving as a cornerstone. By modulating system parameters or applying external pulses, it enables precise manipulation of quantum dynamics, enhancing the performance of tasks like state preparation, gate operations, and qubit readout, all while adhering to platform-specific constraints~\cite{Brif_2010,koch2022quantum, glaser2015training,QOC1,Acin_2018}. 
One particularly powerful subset of quantum control techniques involves shortcuts to adiabaticity (STA) \cite{RevModPhys.91.045001}. However, extending such techniques to dissipative or open quantum systems presents unique challenges due to the interaction between the system and its environment, which induces inevitable decoherence and dissipation {effects}~\cite{breuer2002theory,RevModPhys.89.015001}. These considerations are crucial for realistic quantum technologies, as practical devices often operate in environments. Motivated by these challenges, STA methods have been adapted for open systems to mitigate dissipation and engineer quantum  states under non-unitary dynamics. For example, inverse engineering and counter-diabatic (CD) driving (also known as transitionless quantum driving) have been successfully extended to open systems, facilitating efficient control over dissipative quantum dynamics~\cite{Vacanti_2014,PhysRevLett.122.250402,Alipour2020shortcutsto,PhysRevResearch.2.033178,PhysRevA.104.062421,PhysRevApplied.16.044028,yin2022shortcuts,boubakour2024dynamical}. 

Among the diverse techniques for quantum control, the Pontryagin Maximum Principle (PMP) offers a rigorous mathematical framework for determining optimal solutions in both classical and quantum systems. PMP has been applied to a wide range of quantum tasks, including state preparation~\cite{PhysRevA.97.062343, PhysRevA.74.022306}, physical chemistry~\cite{PhysRevLett.114.233003, PhysRevA.71.063402}, nuclear magnetic resonance~\cite{SKINNER20038, KHANEJA2005296}, quantum metrology~\cite{PhysRevA.105.042621} and quantum information processing~\cite{PRXQuantum.2.017001}. By identifying the optimal trajectory that minimizes a specified cost function under the  constraints, PMP enables the design of precise and efficient quantum control protocols for both closed and open systems~\cite{PRXQuantum.2.030203, PhysRevA.102.052605}.  Furthermore, PMP's capability to provide analytical solutions for shortcuts to isothermality facilitates rapid and energy-efficient thermodynamic processes~\cite{PhysRevLett.128.230603}.  

\begin{figure}[bt]
	\centering
    \includegraphics[width=0.73\linewidth]{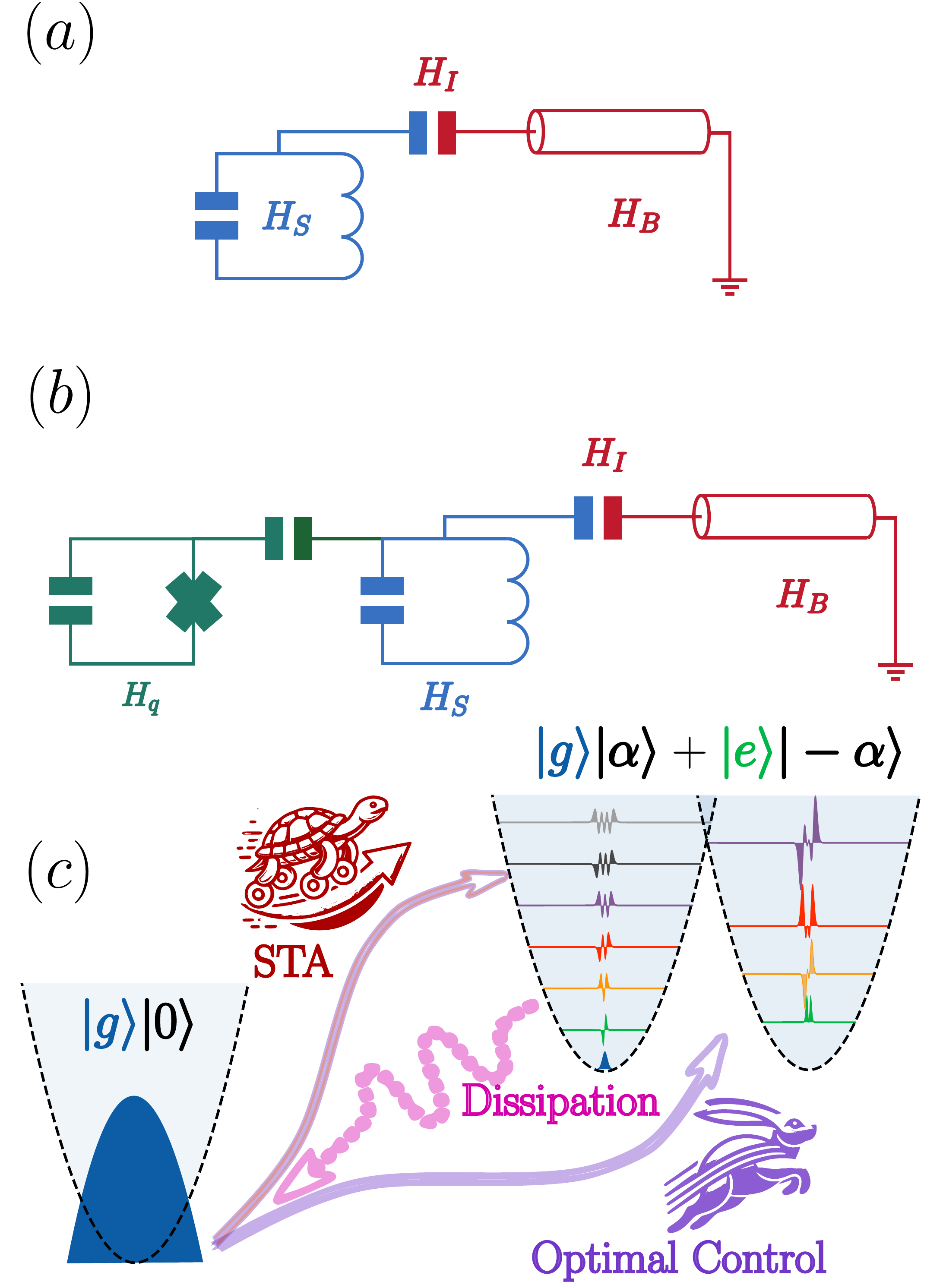}
	\caption{(a) Schematic illustration of driven open quantum system, a LC resonator of frequency $\omega_r$ is coupled to a transmission line resonator mimicking the electronic reservoir. (b) Representation of the cavity-qubit system. (c) Illustration of the optimal control and STA protocols: starting from an eigenstate of the harmonic oscillator, {we transfer this state non-adiabatically to a high-fidelity final coherent state.}}
	\label{Model}
\end{figure}

In this Letter, we present optimized pulses for a driven-dissipative LC circuit, with a focus on circuit quantum electrodynamics ({cQED}) where superconducting qubits couple to microwave resonators ~\cite{RevModPhys.93.025005, Krantz2019}. In cQED, the dissipation resulting from system-environment coupling offers
an ideal platform for studying optimal control in open systems \cite{yin2022shortcuts}.
The optimal control strategies developed here are crucial not only for enhancing the fidelity of quantum state manipulation but also for improving dissipative qubit readout in cQED and, more broadly, in other similar systems involving qubit-cavity coupling \cite{PhysRevLett.119.196802,PhysRevA.98.023849,chessari2024unifying,PhysRevX.14.041023}.

Here, we interpret the system's Langevin equation (LE) as a linear control problem, which allows us to engineer quantum states  with minimal time and low energy consumption. By applying the PMP~\cite{PhysRevApplied.18.054055,ansel2024introduction}, we solve the control problem in the initial stage to obtain analytical pulses capable of transforming the quantum states for a single resonator. {A key preliminary result can be stated as follows:} these specific coherent states are characterized by fixed cavity displacements that minimize energy cost and optimize the time required for a given maximum driving amplitude, in comparison to the adiabatic sinusoidal pulse and its CD assistance~\cite{yin2022shortcuts}. In the second stage, these optimized pulses are used in non-demolition qubit readout across three different regimes: low, intermediate, and high critical photon numbers. Our findings indicate that for large critical photon numbers, the signal-to-noise ratio (SNR) is comparable to that of readout schemes based on longitudinal coupling~\cite{PhysRevLett.115.203601, PhysRevApplied.18.034010}.

\textit{Single-resonator in open system.--}
The circuit in Fig.~\ref{Model}(a) consists of an LC oscillator, capacitively coupled to a transmission line resonator, representing the electromagnetic environment.  The total Hamiltonian  of the open system is given by $H_{\rm{tot}}=H_{S}+H_{B}+H_{I}$ (hereafter $\hbar = 1$),
where the system Hamiltonian  is $H_{S}=\omega_{r} a^{\dagger}a $
with $\omega_{r}$ being the resonator frequency. The bath Hamiltonian is $H_{B}=\int{d\omega' \omega' b^\dagger (\omega')b(\omega')}$, representing a continuum of harmonic oscillators at frequencies $\omega'$. The interaction Hamiltonian is $ H_{I}= i \int{d\omega' \lambda(\omega') [b^\dagger (\omega') a-a^\dagger b(\omega')]}$, describing the coupling between the resonator and the bath, where $\lambda(\omega')$ is the coupling strength. To derive the system dynamics, we trace out the reservoir degrees of freedom, arriving to the LE in the rotating frame~\cite{PhysRevA.31.3761,PhysRevA.46.4363}, $\dot{a}=-i \omega_{r} a-\kappa a/2  -\sqrt{\kappa}a_{\rm{in}}(t)$, where $\kappa = 2 \pi \lambda^2 (\omega')$ is the system decay rate, and $a_{\rm{in}}(t)$ is the input field. By using time-dependent coherent state $\alpha(t)=\left\langle a \right\rangle$ and relating the coherent driving field $\varepsilon(t)$ to the input field as $\varepsilon(t)=\sqrt{\kappa}a_{\rm{in}}(t)$, the LE becomes \cite{SM}
\begin{equation}
\label{da}
\dot{\alpha}=-i \omega_r \alpha-\frac{\kappa}{2} \alpha  -\varepsilon(t),
\end{equation}
where the explicit time dependence of $\alpha (t)$ is omitted for simplicity.
This equation also governs the dynamics of a coherent state in a driven harmonic oscillator in presence of dissipation~\cite{PhysRevApplied.18.054055}, as described by the master equation $d\rho/dt=i[\rho,H'_{S}]+(\kappa/2)(2a\rho a^\dagger-\rho a^\dagger a-a^\dagger a \rho )$~\cite{SM} with $H'_{S}=\omega_r a^\dagger a +i(\varepsilon^*(t)a-\varepsilon(t)a^\dagger)$.
In the following, we outline the method for optimizing $\varepsilon(t)$ to drive the system from an initial state $\alpha(0)$ at $t=0$ to a desired target state $\alpha(t_f)$ at $t=t_f$, while minimizing both the energy cost of the driving field and the required time.

\textit{Minimization of pulse energy and time.--}
Consider the control problem of linear system described by $\dot{\bf{x}}=\hat{A}~{\bf{x}}+\hat{B}~{\bf u}(t)$ \cite{PhysRevApplied.18.054055}, where ${\bf{x}}(t)$ is the system state and ${\bf{u}}(t)$ is the control vector. 
Assuming the matrix $\hat{A}$ is time-independent,  the formal solution is  $\textbf{x}(t)=\exp(\hat{A}t) \textbf{x}(0) + \int^t_{0}\exp[\hat{A}(t-s)]\hat{B}~\textbf{u}(s)ds$, with the initial state $\textbf{x}(0)$. To minimize the functional $J=\int f(\textbf{x},\textbf{u},t)dt$, we incorporate the system's dynamics into the optimization problem through the PMP. Regarding energy optimization, the cost function is $J_E=\int^{t_f}_{0}\textbf{u}^\textbf{T}\textbf{u}dt$. This leads to the Pontryagin Hamiltonian $H_c=\textbf{u}^\textbf{T}\textbf{u}+\textbf{p}^\textbf{T}\cdot \dot{\textbf{x}}$,
where $\textbf{p}$ is the adjoint variable, governed by $\dot{\textbf{p}}=-\partial H_c/\partial \textbf{x}$. For energy minimization, we require $\partial H_c/\partial \textbf{u}=0$, leading to $\dot{\textbf{p}}=-\hat{A}^\textbf{T}\textbf{p}$ and $\textbf{u}=\hat{B}^\textbf{T}\textbf{p}$. Solving these equations for a time-independent $\hat{A}$, we find that the state vector $\textbf{x}(t)$ can be written as
\begin{eqnarray}
    \textbf{x}(t)=e^{\hat{A}t} \textbf{x}(0)+\int^t_0 e^{\hat{A}(t-s)}\hat{B}\hat{B}^\textbf{T}e^{\hat{A}^\textbf{T}(t-s)}ds\textbf{p}(0),
\end{eqnarray}
and the adjoint variable is $\textbf{p}=e^{\hat{A}^\textbf{T}(t_f-t)}\textbf{p}(0)$.
The constant vector $\textbf{p}(0)$ is determined by the boundary condition, $\textbf{p}(0)=[\int^{t_f}_0 \exp[\hat{A}(t_f-s)]\hat{B}\hat{B}^T\exp[\hat{A}^\textbf{T} (t_f-s)]ds]^{-1}[\textbf{x}(t_f)-\exp(\hat{A}t_f)\textbf{x}(0)] $.
Finally, the optimal control is obtained as \cite{SM}
\begin{eqnarray}
\label{optcond}
    \textbf{u}^{{\rm{opt}}}(t)=\hat{B}^\textbf{T}\exp[\hat{A}^\textbf{T}(t_f-t)]\textbf{p}(0).
\end{eqnarray}

\begin{figure}[bt]
\centering
\includegraphics[width=1.0\linewidth]{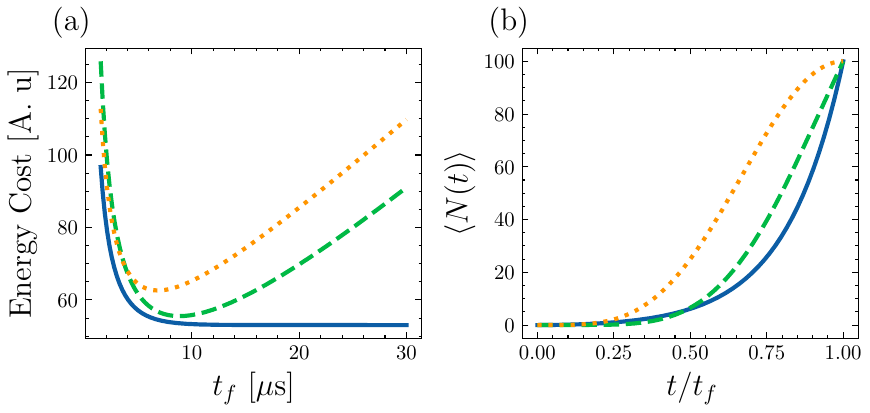}
\caption{(a) Energy cost versus time $t_f$ and (b) time evolution of total photon number for different drivings when $t_f=10~$\textmu s. All schemes, including the energy-optimal (blue solid), the Hahn pulse (orange dotted), and its CD-assisted counterpart (green dashed), achieve the same target state $\alpha(t_f)=10e^{i \vartheta}$ from the initial state $\alpha(0)=0$. Parameters: $\omega_r=2\pi\times 0.3~$MHz and $\kappa=2\pi\times 10~$kHz. The cavity frequency must be chosen to be similar to those used in the optimal driving protocol, since the detuning in the interaction picture needs to be zero. }
\label{drivingEopt}
\end{figure}

Specifically, for the system with coherent state $\alpha(t)= x_1(t)+i x_2(t)$ and driving field  $\varepsilon(t)=\varepsilon_{1}(t)+i\varepsilon_{2}(t)$, substituting these into the control problem (\ref{da}) yields the following system of equations:
\begin{eqnarray}
\label{opt_control_problem}
    \begin{pmatrix}
        \dot{x}_1(t)\\
        \dot{x}_2(t)
    \end{pmatrix}=\begin{pmatrix}
        -\frac{\kappa}{2} & \omega_r\\
        -\omega_r & -\frac{\kappa}{2}
    \end{pmatrix}\begin{pmatrix}
        x_1(t)\\
         x_2(t)
    \end{pmatrix}+\begin{pmatrix}
        -1 & 0\\
        0 & -1
    \end{pmatrix}\begin{pmatrix}
\varepsilon_1(t)\\
\varepsilon_2(t)
\end{pmatrix}.
\end{eqnarray}
By direct comparison, the state vector is $\textbf{x}^\textbf{T}=[ x_1(t), x_2(t)]$, and the control vector is $\textbf{u}^\textbf{T}(t)=[\varepsilon_1(t),\varepsilon_2(t)]$, with the corresponding matrices $\hat{A}$ and $\hat{B}$. Using these, the optimal solution (\ref{optcond}) becomes
\begin{eqnarray}
\label{uopt}
\textbf{u}^{{\rm{opt}}}(t)=\frac{\kappa e^{\kappa(t+t_f)}}{1-e^{\kappa t_f}}\begin{bmatrix}
\alpha(t_f)e^{i\omega_r(t-t_f)}+\alpha^{*}(t_f)e^{-i\omega_r(t-t_f)}\\
i\alpha(t_f)e^{i\omega_r(t-t_f)}-i\alpha^{*}(t_f)e^{-i\omega_r(t-t_f)}
\end{bmatrix},
\end{eqnarray}
which results in the optimal energy cost 
\begin{eqnarray}
\label{Jopt}
    J_E^{\rm{opt}}(t_f)=\frac{4\kappa|\alpha(t_f)|^2}{1-e^{-\kappa t_f}} \xrightarrow{t_f \to \infty} 4\kappa |\alpha(t_f)|^2.
\end{eqnarray}
To assess the reduction in energy cost, we compare the optimized pulses with a typical Hahn (sinusoidal) pulse, defined as $\varepsilon_{{\rm{h}}}(t)=\Omega_{0} \sin^{2}(\pi t/2t_f)$. The constant amplitude $\Omega_{0}$ increases with $t_f$, to satisfy the adiabatic criteria, ensuring that $\alpha(t_f)$ remains $|\alpha(t_f)|e^{i \vartheta}$, with the fixed phase $\vartheta=\pi/2 + \tan^{-1}(\kappa/2\omega_r)$ \cite{SM}. As illustrated in Fig.~\ref{drivingEopt} (a), for small duration ($t_f<10$ \textmu s), both energy costs increase exponentially as the final time $t_f$ decreases, when the state evolves from $\alpha (0)=0$ to $\alpha(t_f) =10e^{i \vartheta}$. However, in the adiabatic limit, $t_f \gg \kappa^{-1} \approx 16$ \textmu s,  $J_E^{{\rm{h}}}(t_f)$ saturates to $  \propto \kappa^2 t_f$, whereas $J_E^{\rm{opt}}(t_f)$ remains constant, $4\kappa |\alpha(t_f)|^2$.  In Fig.~\ref{drivingEopt} (b), both pulses achieve the same final photon number $\langle N(t) \rangle = |\alpha(t)|^2$. However, the optimized pulse keeps the intermediate photon number lower throughout the process, highlighting its advantage in systems with pronounced anharmonicity and ionization \cite{PhysRevX.14.041023}.

\begin{figure}[!t]
\centering
\includegraphics[width=1.0\linewidth]{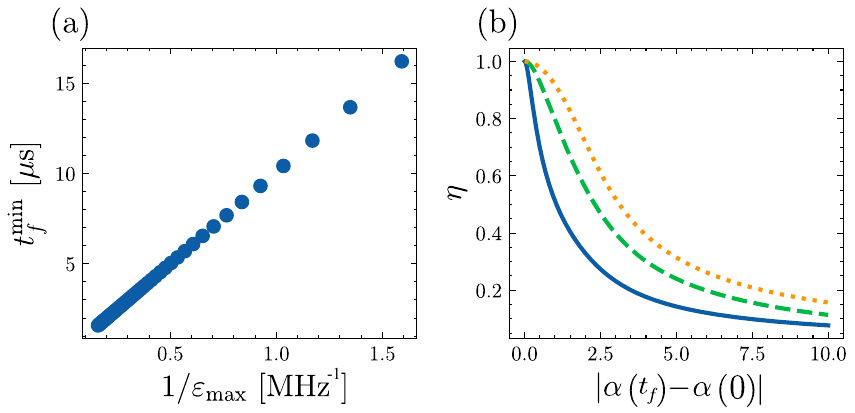}
\caption{(a) Minimal time $t_f^{\rm{min}}$ as a function of $1/\varepsilon_{max}$. (b) Quantum efficiency $\eta$ under different driving schemes: energy-optimal (blue solid) control, time-optimal control (green dashed) and the Hahn pulse assisted by CD assistance (orange dotted). The parameters are the same as those in Fig. \ref{drivingEopt}.}
\label{optime}
\end{figure}

The next step is to find the time-optimal driving field $\varepsilon(t)$ under a constraint, $|\varepsilon(t)|\leq \varepsilon_{max}$. To this end, we parameterize the real and imaginary components of the pulse as $\varepsilon_1(t)=\varepsilon_{max} \cos \phi(t)$ and $\varepsilon_2(t)=\varepsilon_{max} \sin \phi(t)$, where $\phi(t)$ is the phase to be optimized according to the PMP~\cite{pontryagin2018mathematical,PhysRevA.107.032613}. To minimize the cost function $J_T=\int^{t_f}_{0} 1 dt\equiv t_f $, we define the Pontryagin Hamiltonian as $H_c=p_0+\sum p_j\dot{\textbf{x}}_j$, where $\dot{\textbf{x}}_j$ (with $j=1,2$) are given by Eq.~(\ref{opt_control_problem}) and the constant $p_0$ represents an energy offset. The adjoint variable $p_j$ evolves according to the state equation $\dot{p}_j=(-1)^{j+1}\omega_r p_k-\kappa p_j/2$, with $j,k=1,2$ and $k\neq j$. The solution is given by $p_j(t)=e^{-\kappa t/2}[p_{j}(0)\cos(\omega_r t) +(-1)^{j+1}p_{k}(0)\sin(\omega_r t)]$. The energy conservation demands that $e^{\kappa t/2} ({p^2_1}+{p^2_2})={p^2_1}(0)+{p^2_2}(0)=C^2$ ($C$ is a constant), which leads to the initial conditions $p_1(0)=C\cos\theta$ and $p_2(0)=C\sin\theta$. According to PMP, optimizing $H_c$ requires $\partial H_{c}/\partial \phi=0$, which yields the condition $\tan \phi^{{\rm{opt}}}(t)=p_2(t)/p_1(t)$, or equivalently, $\phi^{\rm{opt}}(t) = \omega_r t + \theta$. Solving the LE in Eq.~(\ref{da}) using $\phi^{\rm{opt}}(t) $, we get
\begin{equation}
\label{solfield}
    \alpha^{\rm{opt}}(t)=\alpha(0)~e^{-(\frac{\kappa}{2}+i\omega_r)t}+\frac{2\varepsilon_{\rm{max}}~e^{i\theta}}{\kappa+4i\omega_r}\left[e^{-(\frac{\kappa}{2}+i\omega_r)t}-1\right].
\end{equation}
The optimal time $t_f^{\rm{min}}$ is determined by minimizing $|\alpha^{\rm{opt}}(t_f^{\rm{min}})|-|\alpha(t_f)|$. The phase $\theta$ is then adjusted to match the phase of $\alpha^{{\rm{opt}}}(t^{\rm {min}}_f)$ with that of the target state $\alpha(t_f)$. By satisfying these conditions, the system reaches the target state, e.g., $\alpha(t_f)= 10 e^{i \vartheta}$, in the shortest possible time. Fig.~\ref{optime}(a) shows the minimal time as a function of  $1/\varepsilon_{max}$, indicating an exponential decrease with increasing driving amplitude,  e.g. $t_f^{\rm{min}}=1.01~$\textmu s when $\varepsilon_{\rm{max}}=10 $~MHz.

To further compare the energy- and time-optimized pulses, the CD driving is given by~\cite{yin2022shortcuts}
\begin{equation}
\label{eqcd}
\varepsilon_{CD}(t)=\varepsilon(t)-i\frac{\dot{\varepsilon}(t)}{\omega_r-i\kappa/2},
\end{equation} 
where the adiabatic reference $\varepsilon(t)$ can be the Hahn pulse $\varepsilon_{{\rm{h}}}(t)=\Omega_{0} \sin^{2}(\pi t/2t_f)$ with a fixed $\Omega_0 = |\alpha(t_f)| \sqrt{\omega^2_r +\kappa^2 / 4}$ \cite{SM}. Physically, this CD protocol can be implemented by driving the orthogonal quadrature resonator. In other words, for voltage-controlled pulses, the CD terms should be a current-controlled one, and vice versa. The inclusion of additional term guarantees to reach the equilibrium state without introducing the excited photons. Fig.~\ref{drivingEopt}(a) shows that,  while the CD pulse is more energy-efficient than the Hahn pulse, it is still less efficient than the energy-optimal pulse. In the adiabatic limit, the CD pulse exhibits the same slope for the energy cost as the Hahn pulse, but with a constant residual energy.
 
To quantify the quantum speed limits \cite{Deffner_2017}, we invoke the Mandelstam-Tamm bound \cite{mandelstam1991uncertainty}: $ \int_{0}^{t_f} \Delta H dt \geq S_0$, where the geodesic distance $S_0 =\arccos(|\langle\psi (0) |\psi (t_f)\rangle|)$ between the initial and final states is determined by 
 the Fubini-Study metric, and the standard deviation of the energy, $ \Delta H = (\langle H^2 \rangle-\langle H \rangle^2)^{1/2}$, quantifies the energy uncertainty during the evolution. For our control problem involving coherent states, the effective Hamiltonian is given by  $H= i (\dot{\alpha} a^{\dagger} -\dot{\alpha}^{*} a)$, and the quantum efficiency is expressed as \cite{yin2022shortcuts}:
 \begin{equation}
 \label{quantum efficiency}
 \eta =\frac{S_0}{\int_{0}^{t_f} \Delta H dt}  \leq \frac{ \arccos{(e^{-|\alpha(t_f)-\alpha(0)|^2/2})}}{ | \alpha(t_f)-\alpha(0)|},
 \end{equation} 
with the energy uncertainty being $\Delta H = |\dot{\alpha}(t)|$. As illustrated in Fig.~\ref{optime}(b),  
the CD driving achieves the theoretical upper bound of efficiency, as the energy uncertainty precisely matches the additional term $|\varepsilon_{CD}(t)-\varepsilon(t)|$. In contrast, our optimal controls approach, but do not fully reach this bound, due to the constraint imposed on $\varepsilon(t)$, rather than $ |\dot{\alpha}(t)|$.


\textit{Qubit-resonator interaction.--}
We now extend the control framework to a system consisting of an LC resonator dispersively coupled to a qubit, as depicted in Fig.~\ref{Model}(b). In the dispersive approximation, the Jaynes-Cumming Hamiltonian is given by \cite{PhysRevApplied.22.014079}
\begin{eqnarray}
\label{Hread}
H_S=\chi\sigma_z a^\dagger a +\left(\frac{\omega_q+\chi}{2} \right) \sigma_z,
\end{eqnarray}
where $\omega_q$ is the qubit frequency, and $\chi=g^2/(\omega_q-\omega_r)$ is the effective Stark-shift, moreover, $\sigma_z$ is the Pauli-$z$ operator. This Hamiltonian represents the conditional shift on the resonator frequency due to the qubit and vice versa, which can be used for qubit readout when the system is driven by an external field $\varepsilon(t)$ in LE~\cite{PhysRevA.69.062320}. For better readout contrast, larger driving amplitudes are needed to separate the qubit states more effectively on the resonator's IQ plane. However, such amplitudes cannot exceed the fundamental limits imposed by the dispersive approximation that sets an upper bound on the photon number $\bar{n}_{\rm{crit}}=(\omega_q-\omega_r)^2/4g^2$~\cite{PhysRevA.76.042319}.
Moreover, the drive must be chosen carefully to avoid ionizing the artificial atom~\cite{PhysRevApplied.18.034031}.  From the perspective of optimal control, it is essential to design the driving field that maximize both readout contrast and energy efficiency while ensuring the dispersive approximation holds and the artificial atom is not ionized.

\begin{figure}[!t]
\centering
\includegraphics[width=1.0\linewidth]{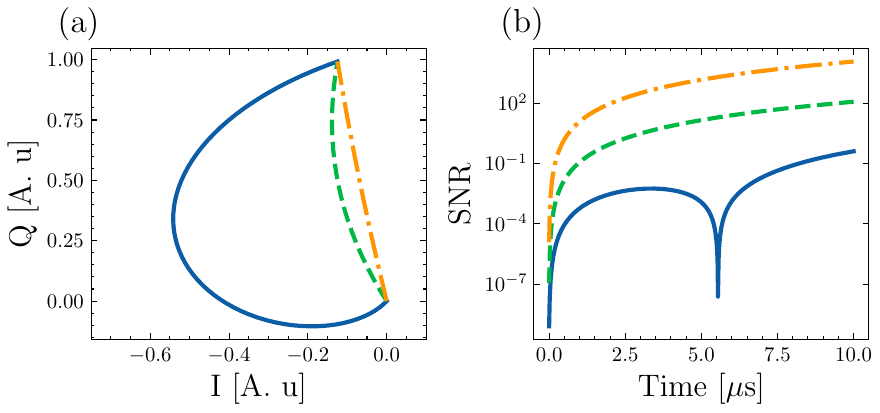}
\caption{(a) The evolution in the IQ plane (normalized by $\max\{\sqrt{Q^2+I^2}\}$) for a qubit-cavity system initialized in the state $\ket{0,g}$ with an initial $\langle N(0) \rangle=0$. (b) SNR versus  measurement time for the same  critical photon number, with $\varphi=0$. Here the critical photon numbers: $\bar{n}_{\rm{crit}}=1$  (solid blue),  $\bar{n}_{\rm{crit}}=10$ (green dashed), and $\bar{n}_{\rm{crit}}=100$ (orange dot-dashed) are chosen, corresponding to different final $\langle N(t_f) \rangle$.  Parameters:  $t_f=10~$\textmu s, $\omega_r=2\pi\times 6~$GHz, $\omega_q=2\pi\times 4~$GHz, $\kappa=2\pi\times 10~$kHz, and the corresponding qubit-cavity coupling strengths of  $g=2\pi\times(1000,10,1)~$MHz.}
\label{disSNR}
\end{figure}

We focus on system parameters resulting in three different critical photon numbers $\bar{n}_{\rm{crit}}=(1, 10, 100)$, which define the final boundary conditions for $\langle N(t_f)\rangle \equiv |\alpha(t_f)|^2=\bar{n}_{\rm{crit}}$, i.e., $\alpha(t_f)=(1,\sqrt{10},10)e^{i \vartheta}$. To maintain the validity of the dispersive approximation, we set the qubit-cavity detuning to $|\omega_q-\omega_r|= 2\pi \times 2~$GHz. Moreover, to engineer optimized readout pulses for the qubit-cavity system, we note that the Hamiltonian~(\ref{Hread}) shares the same LE (\ref{da}), with the substitution $\omega_r\rightarrow \chi_z$, where $\chi_z=\pm\chi$ is state-dependent Stark-shift~\cite{PhysRevA.69.062320,PhysRevLett.119.196802}. Consequently, we can obtain the optimal control for energy minimization using Eq.~(\ref{uopt}), which gives:
\begin{eqnarray}
\textbf{u}^{\rm{opt}}(t)=\frac{\kappa e^{\kappa(t+t_f)}}{1-e^{\kappa t_f}}\begin{bmatrix}
\alpha(t_f)e^{i\chi_z(t-t_f)}+\alpha^{*}(t_f)e^{-i\chi_z(t-t_f)}\\
i\alpha(t_f)e^{i\chi_z(t-t_f)}-i\alpha^{*}(t_f)e^{-i\chi_z(t-t_f)}
\end{bmatrix}.
\label{optenergy}
\end{eqnarray}
Similarly, we estimate the minimal control time $t_f^{\rm min}$ under the constraint of maximal driving amplitude that ramps the resonator with photon numbers near $\bar{n}_{\rm{crit}}$ using Eq.~(\ref{solfield})
\begin{eqnarray}
\label{opttime}
    \alpha^{\rm{opt}}(t)&=&e^{-(\frac{\kappa}{2}+i\chi_z)t}+\frac{2\varepsilon_{\rm{max}}e^{i\theta} }{\kappa+4i\chi_z}\left[e^{-(\frac{\kappa}{2}+i\chi_z)t}-1\right].
\end{eqnarray}
The primary difference from the single-resonator case is that the field displacement now depends on the qubit state. A larger separation between the qubit states leads to better readout outcome. 
Fig.~\ref{disSNR}(a) shows the normalized IQ plane of the resonator, corresponding to the real and imaginary parts of $\alpha^{\rm{opt}}(t) $, when we initialize the qubit in the ground state $\ket{g}$, with a readout time of $t_f=10~$\textmu s. Since the time is fixed, we apply the optimal protocol (\ref{optenergy}) for energy minimization in the subsequent calculations with $\chi_z =\chi$ corresponding to the solution for $\ket{e}$. The result for $\chi_z =-\chi$ is equivalent as it only changes the direction in which the state $\ket{g}$ displaces in the resonator phase space. The optimal pulse for smaller $\bar{n}_{\rm{crit}}$ shows an exotic trajectory on the resonator IQ plane. This deviation arises due to the large coupling strength $g=2\pi\times1.0~$GHz, which is considerably higher than typical cQED implementations~\cite{RevModPhys.93.025005}. As a result,  nonlinearities, such as Kerr effect, dominate the dynamics~\cite{PhysRevLett.93.207002,PhysRevLett.105.173601} displacing the resonator state. The competition between these displacements modifies substantially the expected IQ trajectory. However, as the coupling strength decreases, the Kerr nonlinearities are suppressed, and the linear ramping becomes the dominant effect, as illustrated by the orange dot-dashed line in Fig.~\ref{disSNR}(a).

Another way to evaluate the performance of optimal pulses is by assessing the quality of the readout process using SNR. 
The SNR is defined as the ratio between the homodyne signal and its fluctuations. Specifically, the homodyne signal is given by $|\langle\hat{\mathcal{M}}_{\ket{e}}\rangle - \langle\hat{\mathcal{M}}_{\ket{g}}\rangle|$, where  the average homodyne signal for the state $\ket{\ell} \in \{\ket{g}, \ket{e}\}$ is defined as
$\hat{\mathcal{M}}_{\ket{\ell}}(\tau)=\sqrt{\kappa}\int_{0}^{\tau}dt [a^{\dag}_{\rm{out}}(t)\exp(i\varphi) + a_{\rm{out}}(t)\exp(-i\varphi)]$, with $a_{\rm{out}}(t)$ being the output field, determined by the input-output relation $a_{\rm{out}}(t) = a_{\rm{in}}(t) + \sqrt{\kappa}a$. Thus, the SNR is given by  
\begin{eqnarray}
\label{SNR}
{\rm{SNR}} = \frac{|\langle\hat{\mathcal{M}}_{\ket{e}}\rangle - \langle\hat{\mathcal{M}}_{\ket{g}}\rangle|}{\sqrt{\langle\hat{\mathcal{M}}_{{\rm{N}}\ket{e}}^{2}\rangle + \langle\hat{\mathcal{M}}_{{\rm{N}}\ket{g}}^{2}\rangle}},
\end{eqnarray}
where  the fluctuations from the noise homodyne operator is $\hat{\mathcal{M}}_{{\rm{N}}\ket{\ell}}=\hat{\mathcal{M}}_{\ket{\ell}}-\langle\hat{\mathcal{M}}_{\ket{\ell}}\rangle$. 
We illustrate the SNR behavior for three different critical photon numbers in Fig.~\ref{disSNR}(b), with $\varphi=0$. Similar to the IQ trajectory, the case $\bar{n}_{\rm{crit}}=1$ exhibits exotic behavior. This case has the lowest SNR, as it is challenging to drive the resonator close to its critical value without significant pullback from the Kerr correction. As a result, the contrast between the states is not optimal, and noise dominates the signal. Additionally, the SNR evolution displays two minima, which stem from the exotic trajectory in the IQ plane: as the resonator states pass through origin multiple times, coherence is lost,  causing the noise accumulation and a consequent SNR reduction. More details on time-optimal protocol (\ref{opttime}) and CD driving are available in Ref. \cite{SM}. We demonstrate that energy-optimal control achieves higher SNR faster as compared to other schemes, and SNR under time-optimal driving reaches its maximum when $\varphi= n \pi$, where $n \in \mathbb{Z}$.

 
\begin{figure}[!t]
\centering
\includegraphics[width=1.0\linewidth]{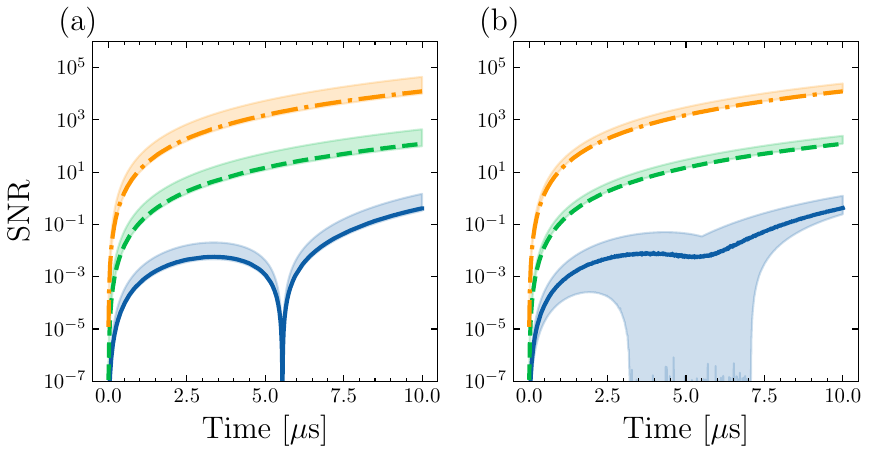}
\caption{(a) Resonator frequency and (b) qubit frequency mismatch on the SNR for different critical photon numbers:  $\bar{n}_{\rm{crit}}=1$ (blue solid), $\bar{n}_{\rm{crit}}=10$ (green dashed), and  $\bar{n}_{\rm{crit}}=100$ (orange dot-dashed). Shaded areas indicate the variance of the averaged SNR across different realizations.
The parameters are the same as those in Fig.~\ref{disSNR}.}
\label{SNRrob}
\end{figure}

For larger $\bar{n}_{\rm{crit}}$, our optimized pulse exhibits a behavior comparable to cQED readout based on longitudinal coupling, as discussed in recent works~\cite{PhysRevLett.115.203601,PhysRevApplied.18.034010}. In these cases, the SNR is higher on a timescale shorter than the decay rate of the resonator.  Moreover, the maximal peak of SNR increases monotonically  with the critical photon number. This trend can be attributed to the fact that, as the photon number increases, the displacement in the resonator also becomes larger, thereby enhancing the contrast and facilitating better discrimination between qubit states.  Additionally, it is possible to further improve SNR for these optimal pulses by initializing the resonator in a squeezed state~\cite{PhysRevLett.115.203601}. 

To further evaluate the robustness of the optimized pulses, we analyze the SNR under 20$\%$ frequency errors~\cite{Krantz2019} in both resonator and qubit subsystems across 1000 samples, as depicted in Fig.~\ref{SNRrob}.
For the resonator,  frequency variations introduce oscillatory contribution in Eq.~(\ref{da}), affecting the pointer state's trajectory yielding smaller SNR, whereas the mismatches on the qubit frequency modifies the dispersive approximation and consequently the critical photon number $\bar{n}_{\rm{crit}}$. The performance under typical parameter fluctuations shows that the energy-optimal pulses are more sensitive to errors at small $\bar{n}_{\rm{crit}}$, as these mismatches significantly alters $\bar{n}_{\rm{crit}}$, increasing the likelihood that the pulse drives the system beyond the dispersive approximation. In contrast, for larger critical photon numbers, the pulses demonstrate the robustness against parameter variations.

\textit{Conclusion.--}
In summary, we applied PMP to design time- and energy-optimised pulses for state transformation and non-demolition qubit readout in open quantum systems within cQED. Using the LE, we derived analytical pulses that respect various constraints. These optimized pulses efficiently engineer the displaced light states with fixed cavity displacement and photon number, exhibiting exponential scaling of energy cost with the operation time. The time-optimized pulse achieves microsecond-scale duration at MHz driving amplitudes. For high-fidelity readout, in dispersively coupled resonator-qubit systems, our optimal pulses achieve nearly unit SNR within a timescale shorter than the resonator's decay rate
for large critical photon numbers.

Our work paves the way for similar pulse optimization in other platforms, such as quantum dots \cite{PhysRevB.100.245427,PhysRevLett.129.066801,PhysRevApplied.20.064005}. Future research could explore numerical optimization  \cite{gautier2024optimal} or machine learning \cite{rinaldi2021dispersive,PhysRevX.12.031017,PRXQuantum.2.040355,Ding_2023} to improve robustness against errors and noise, addressing more complex optimization problems involving decoherence and anharmonicity. Yet, this study not only explores fundamental speed limits \cite{Funo_2019,PhysRevLett.110.050403} in quantum open systems, but also supports practical applications in dissipative qubit readout within cQED.

\textit{Acknowledgement.--}
We are grateful to Sigmund Kohler, 
Ricardo Puebla and Shuoming An for their valuable discussions.
This work is supported by NSFC (12075145 and  12211540002), STCSM (2019SHZDZX01-ZX04), the Innovation Program for Quantum Science and Technology (2021ZD0302302),  HORIZON-CL4-2022-QUANTUM-01-SGA Project No. 101113946 OpenSuperQPlus100 of the EU Flagship on Quantum Technologies, the project grant PID2021-126273NB-I00 funded by MCIN/AEI/10.13039/501100011033 and by ``ERDF A way of making Europe" and ``ERDF Invest in your Future", the Spanish Ministry of Economic Affairs and Digital Transformation through the QUANTUM ENIA project call-Quantum Spain project.
F.A.C.L. thanks to the German Ministry for Education and Research, under QSolid, Grant no. 13N16149.

\nocite{*}
\bibliography{ref}

\begin{figure*}
\includegraphics[page=1,width=1.0\linewidth]{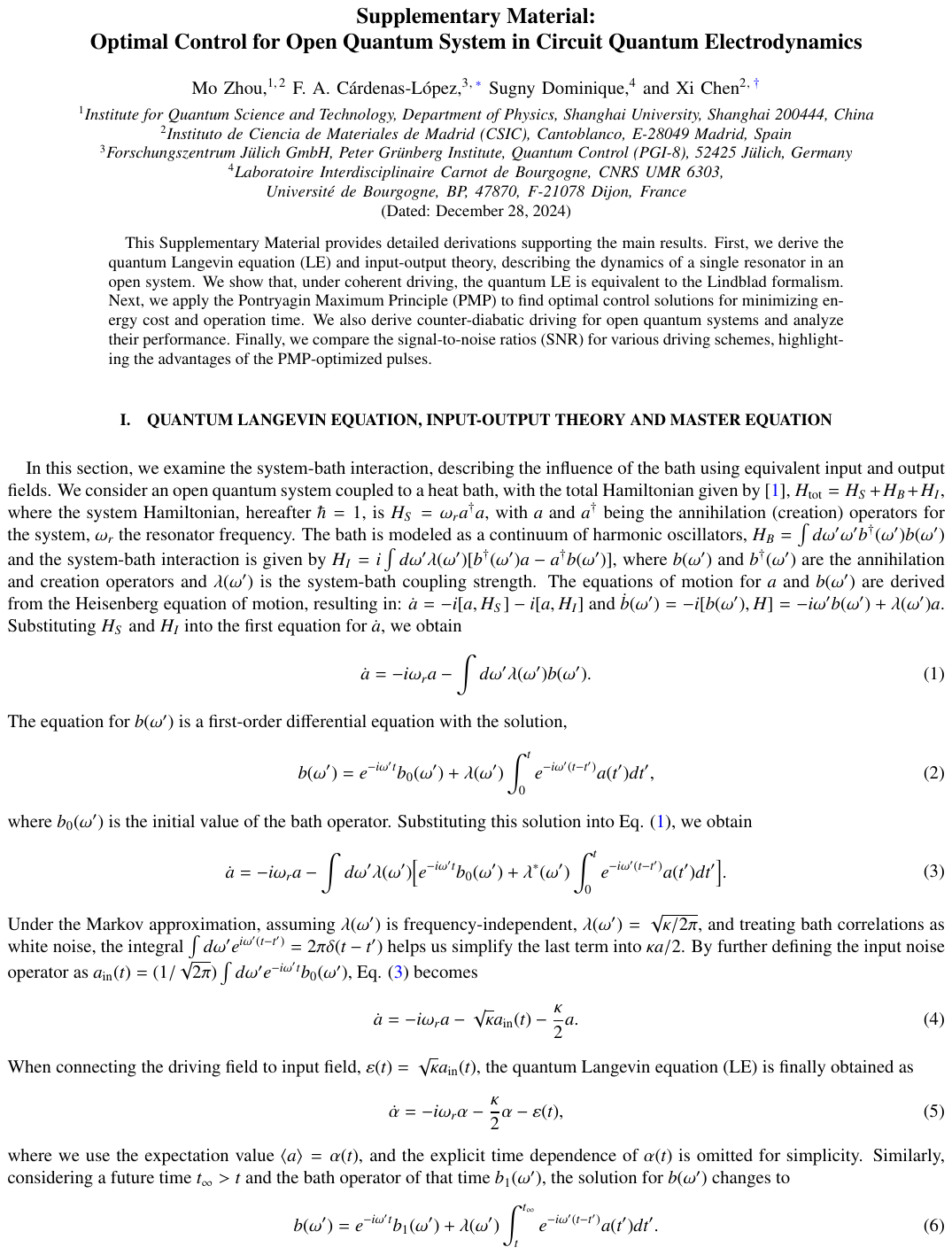}
\end{figure*}
\begin{figure*}
	\includegraphics[page=2,width=1.0\linewidth]{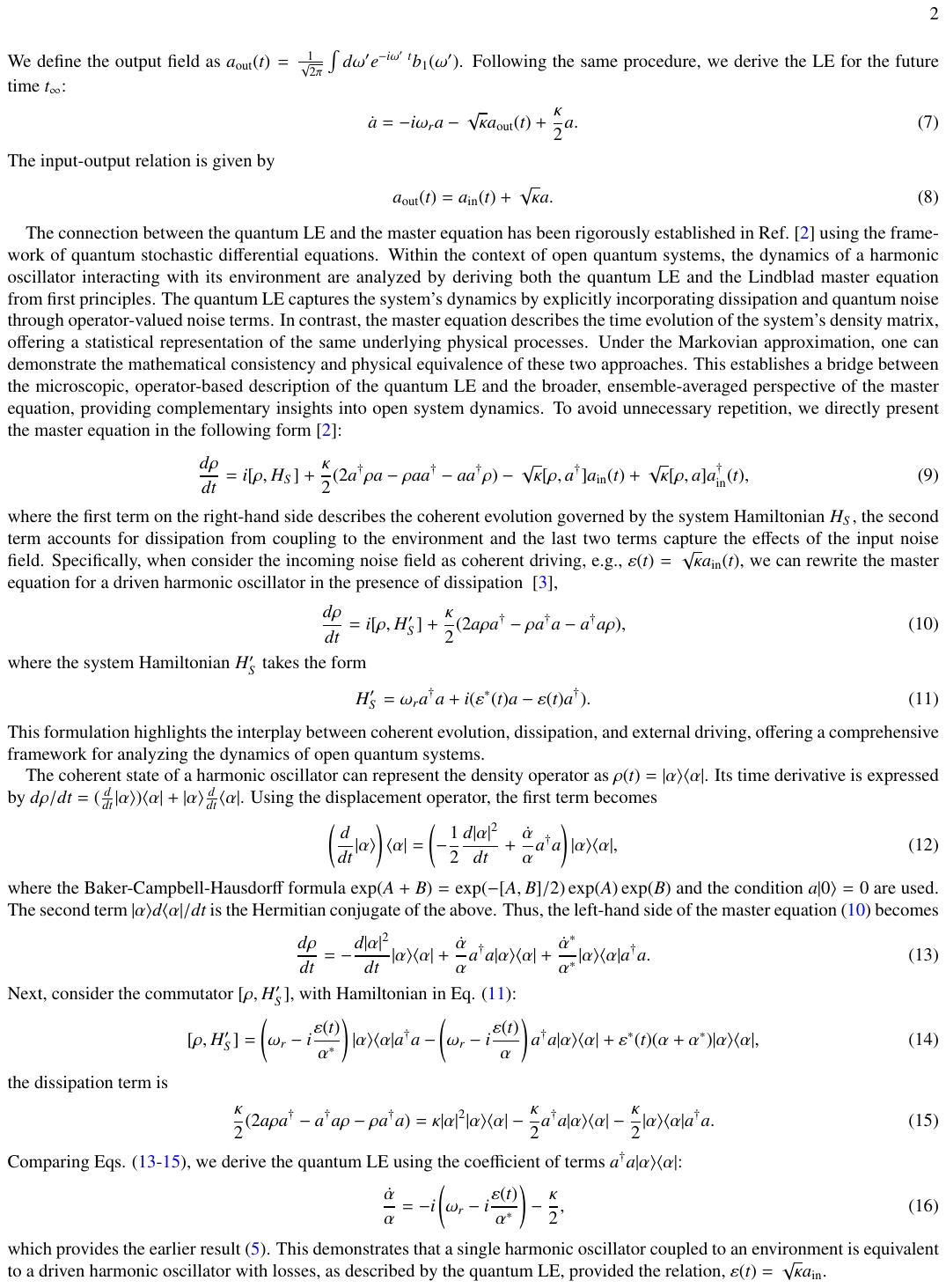}
\end{figure*}
\begin{figure*}
	\includegraphics[page=3,width=1.0\linewidth]{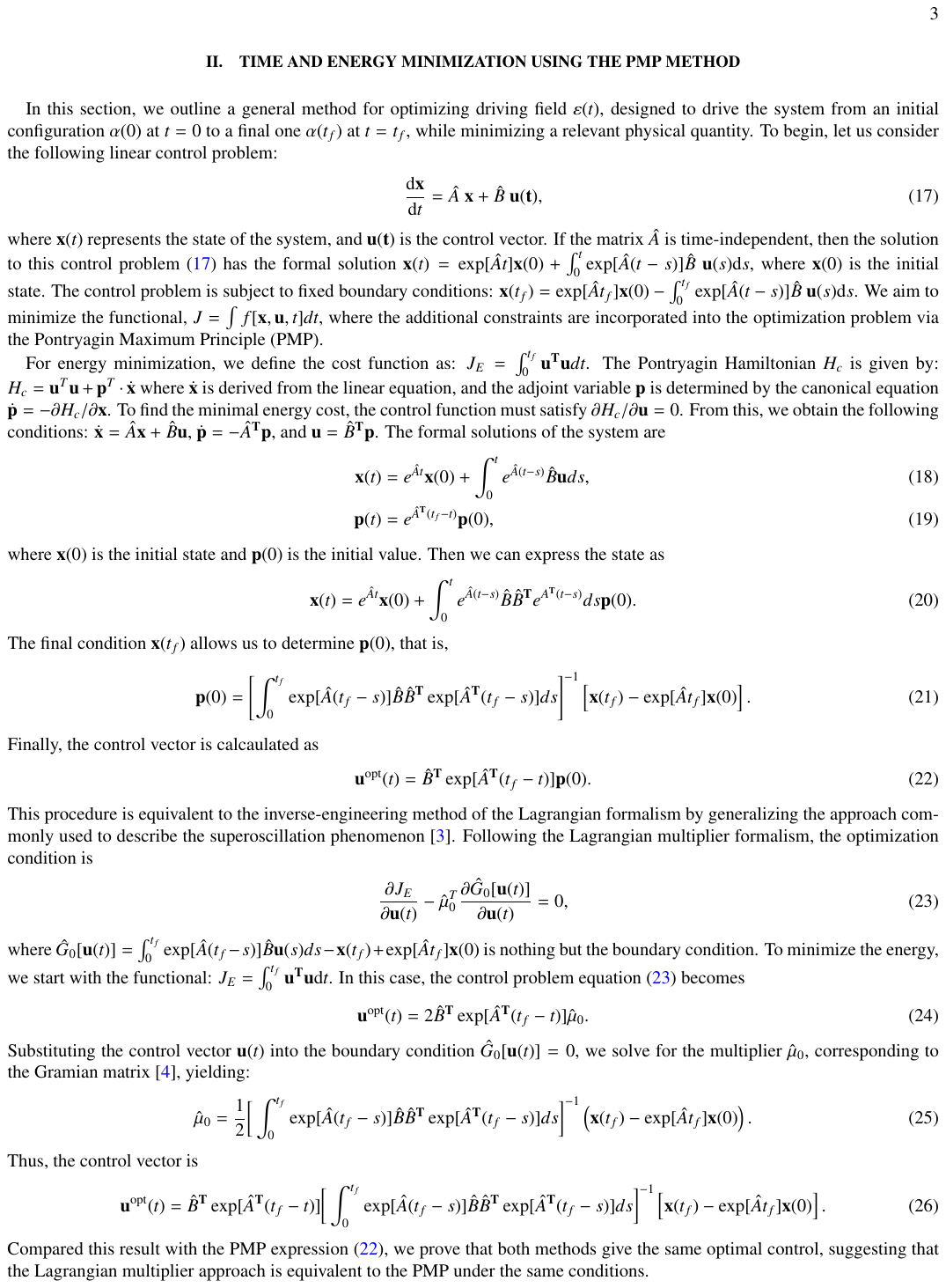}
\end{figure*}
\begin{figure*}
	\includegraphics[page=4,width=1.0\linewidth]{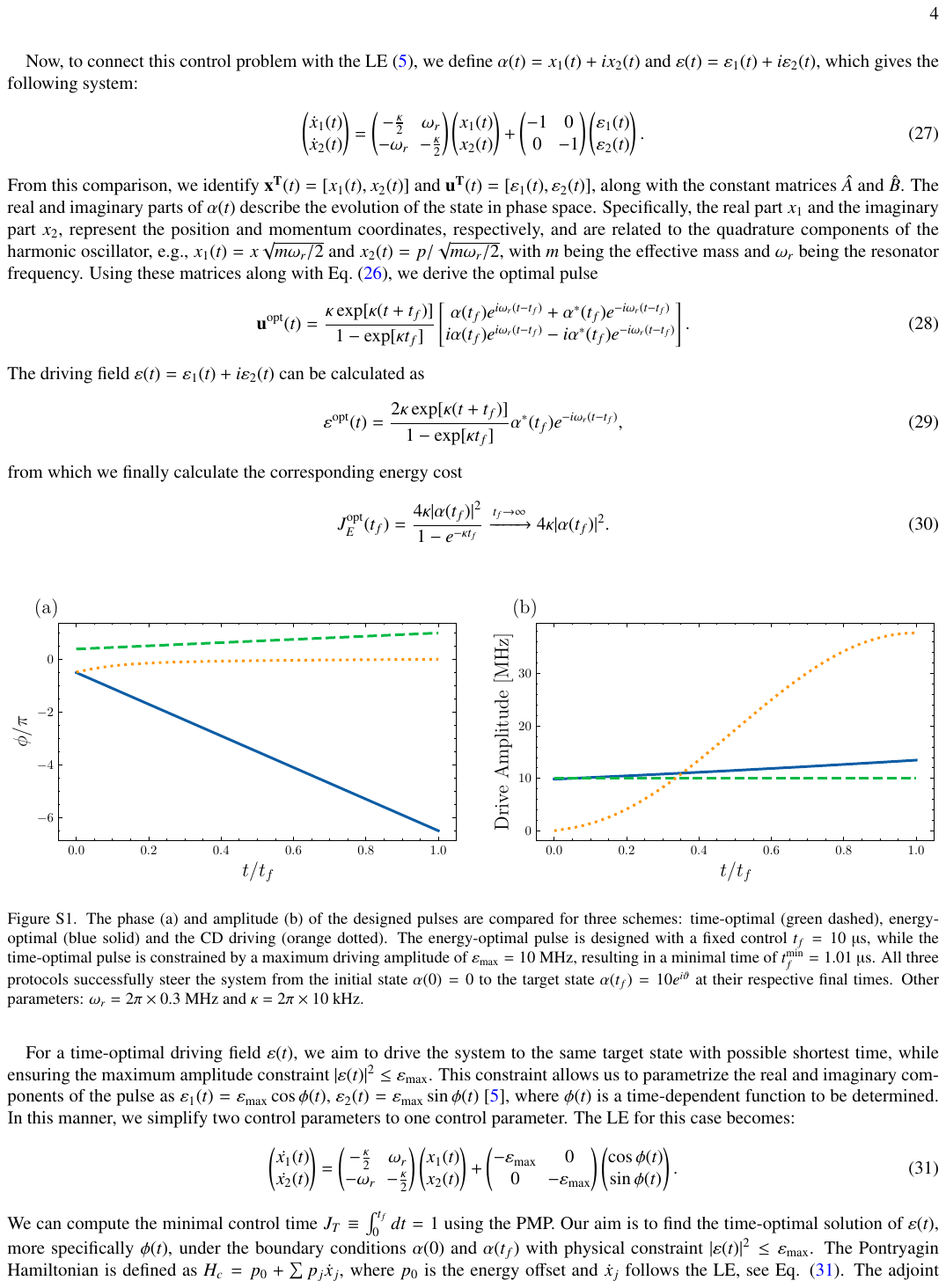}
\end{figure*}
\begin{figure*}
	\includegraphics[page=5,width=1.0\linewidth]{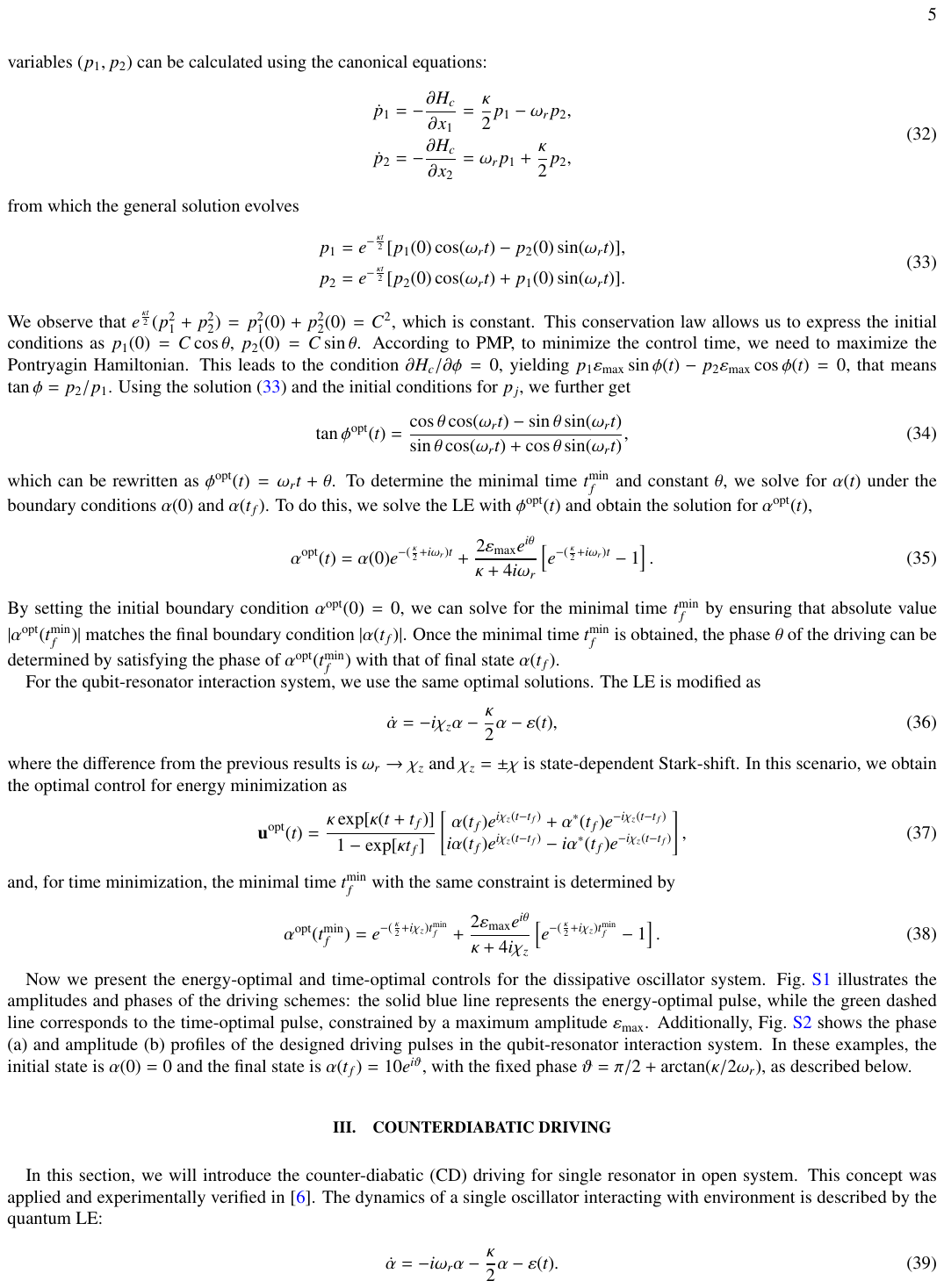}
\end{figure*}
\begin{figure*}
	\includegraphics[page=6,width=1.0\linewidth]{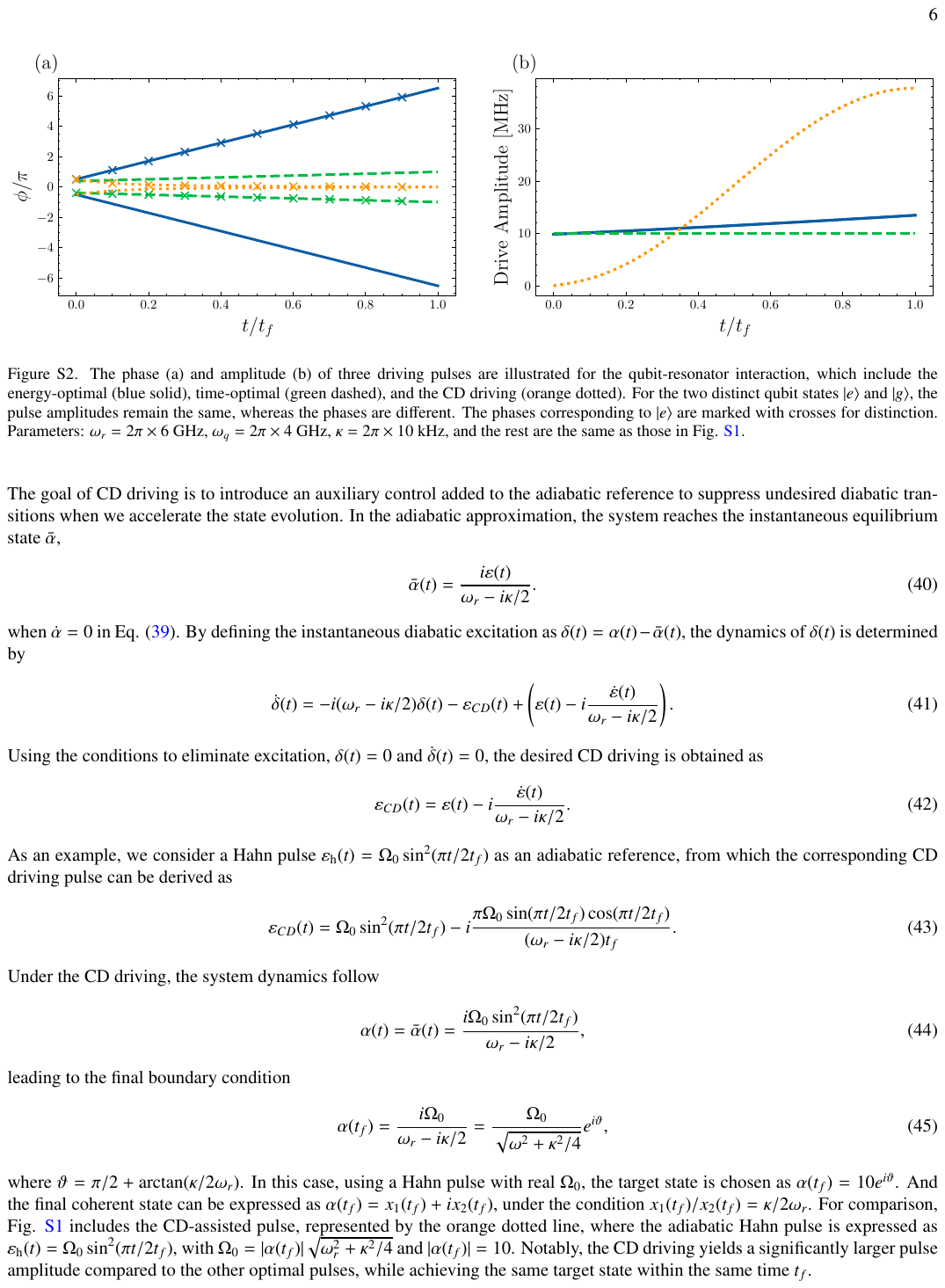}
\end{figure*}
\begin{figure*}
\includegraphics[page=7,width=1.0\linewidth]{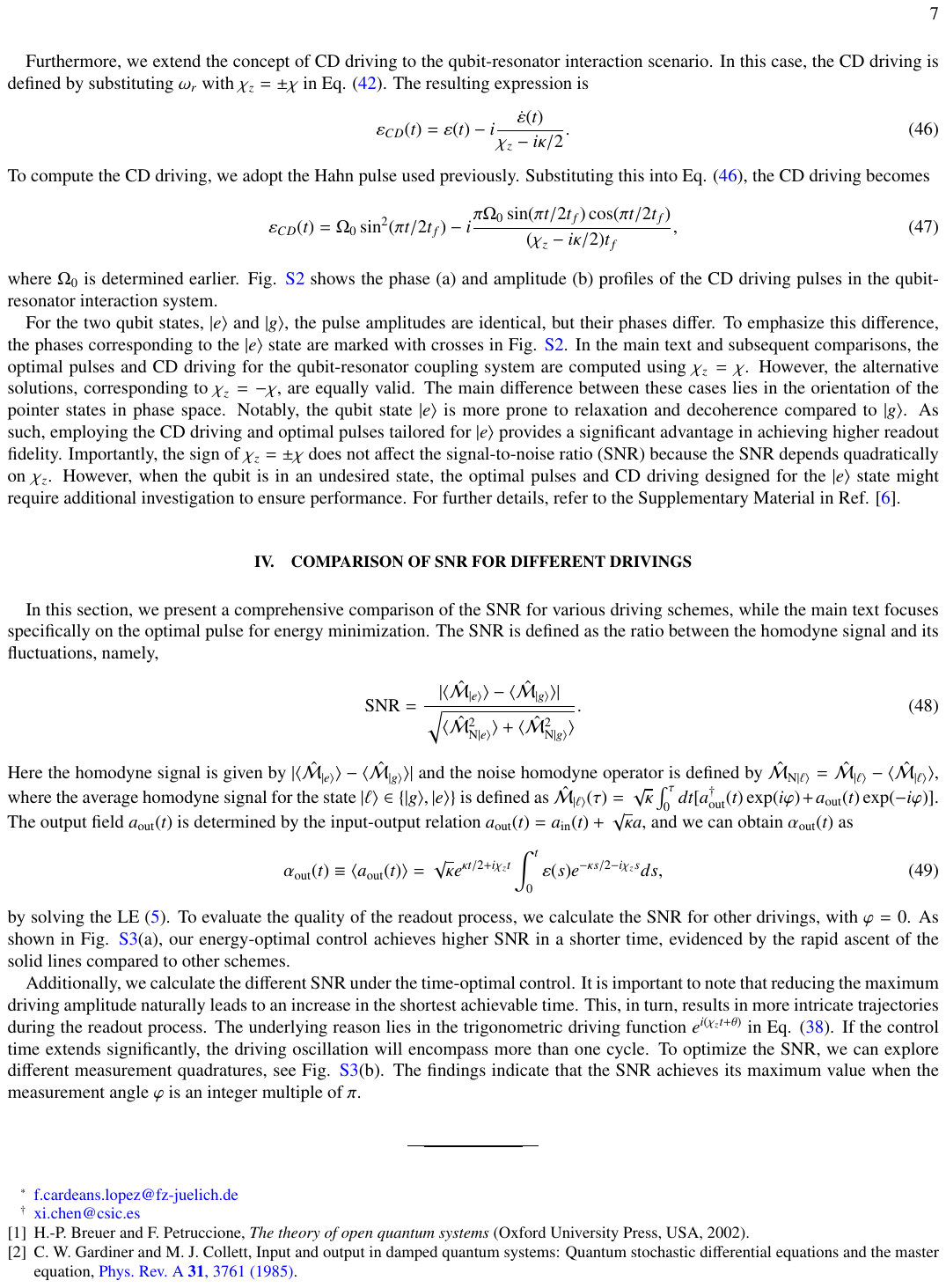}
\end{figure*}
\begin{figure*}
\includegraphics[page=8,width=1.0\linewidth]{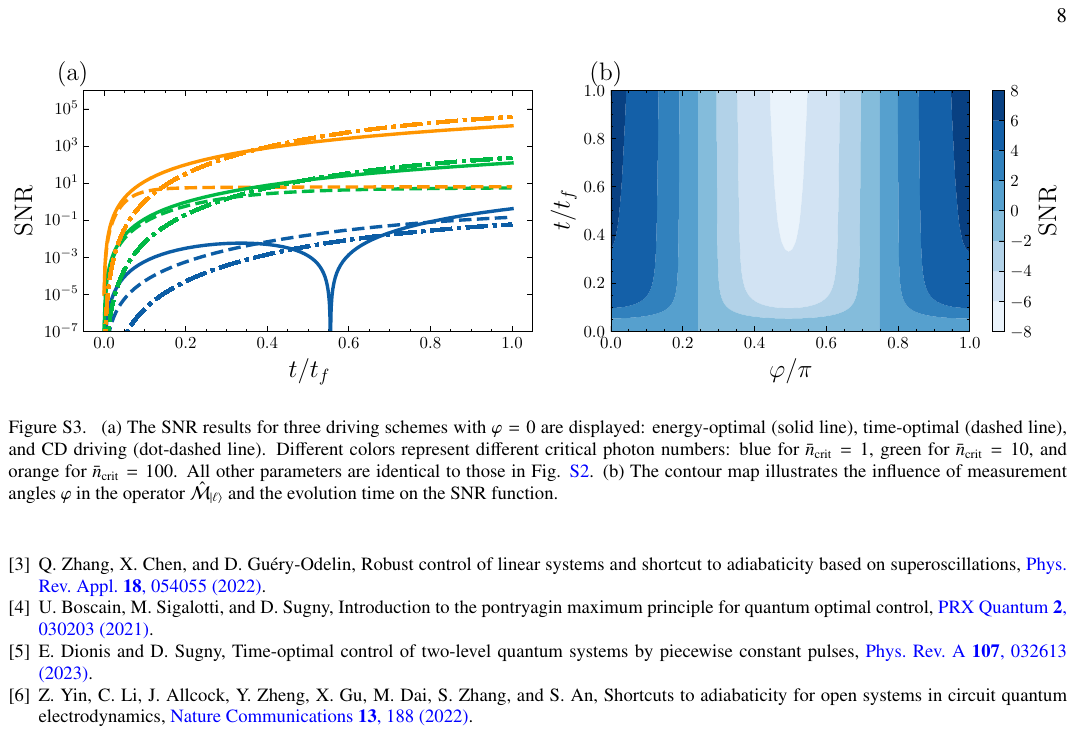}
\end{figure*}

\end{document}